\documentstyle[12pt,epsf]{article}
\renewcommand{\bar}[1]{\overline{#1}}

\begin{document}
\begin{flushright}
USM-TH-87\\
CPT-2000/P.3952
\end{flushright}
\bigskip\bigskip

\centerline{\large \bf $\Lambda$, $\bar{\Lambda}$ polarization and
spin transfer in lepton deep-inelastic scattering}

\vspace{22pt}
\centerline{\bf 
Bo-Qiang Ma\footnote{e-mail: mabq@phy.pku.edu.cn}$^{a}$,
Ivan Schmidt\footnote{e-mail: ischmidt@fis.utfsm.cl}$^{b}$,
Jacques Soffer\footnote{e-mail: Jacques.Soffer@cpt.univ-mrs.fr}$^{c}$,
Jian-Jun Yang\footnote{e-mail: jjyang@fis.utfsm.cl}$^{b,d}$}

\vspace{8pt}

{\centerline {$^{a}$Department of Physics, Peking University,
Beijing 100871, China,}}

{\centerline {CCAST (World Laboratory),
P.O.~Box 8730, Beijing 100080, China,}}

{\centerline {and Institute of High Energy Physics, Academia
Sinica, P.~O.~Box 918(4),}

{\centerline {Beijing 100039, China}

{\centerline {$^{b}$Departamento de F\'\i sica, Universidad
T\'ecnica Federico Santa Mar\'\i a,}}

{\centerline {Casilla 110-V, 
Valpara\'\i so, Chile\footnote{Mailing address}}

{\centerline {$^{c}$Centre de Physique Th$\acute{\rm{e}}$orique,
CNRS, Luminy Case 907, F-13288 Marseille Cedex 9, France}}

{\centerline {$^{d}$Department of Physics, Nanjing Normal
University,}}

{\centerline {Nanjing 210097, China}}

\vspace{10pt}
\begin{center} {\large \bf Abstract}

\end{center}
The flavor and helicity distributions of the $\Lambda$ and
$\bar{\Lambda}$ hyperons for both valence and sea quarks are
calculated in a perturbative QCD (pQCD) based model. We relate
these quark distributions to the fragmentation functions of the
$\Lambda$ and $\bar{\Lambda}$, and calculate the $z$-dependence
of the longitudinal spin transfer to the
$\Lambda$ and $\bar{\Lambda}$ in lepton deep-inelastic scattering
(DIS). It is shown that the spin transfer to the $\Lambda$ is
compatible with the first HERMES results at DESY and further tests
are suggested. We also make predictions for the $z$-dependence
of the $\Lambda$ and $\bar{\Lambda}$ longitudinal
polarizations in neutrino (antineutrino) DIS processes. We
investigate the sea contribution to the fragmentation functions,
and we test a possible scenario where sea quarks in $\Lambda$ (or
sea antiquarks in $\bar{\Lambda}$) are negatively polarized,
whereas sea antiquarks in the $\Lambda$ (or sea quarks in
$\bar{\Lambda}$ ) are positively polarized. The asymmetry of the
polarized fragmentation functions of sea quarks and antiquarks to
$\Lambda$ and $\bar{\Lambda}$ provides a way to understand the
different behaviour between the $\Lambda$ and $\bar{\Lambda}$ spin
transfers observed in the recent E665 experiment at FNAL.

\vfill
\centerline{PACS numbers: 14.20.Jn, 12.38.Bx, 13.87.Fh, 13.88.+e}
\vfill
\centerline{To be published in Eur. Phys. J. C}
\vfill

\newpage
\section{Introduction}

The fragmentations of the $\Lambda$ hyperon, in particular for
polarized $\Lambda$, have received a lot of attention recently,
both theoretically and experimentally [1-23].
There are several reasons for this. First, the spin structure of
the $\Lambda$ is rather simple in the naive quark model, since the
spin of the $\Lambda$ is completely carried by the valence strange
quark, while the up and down quarks form a system in a spin
singlet state and give no contribution to the $\Lambda$ spin.
Therefore any observation of polarization of the up and down
quarks in a $\Lambda$, which departs from this simple picture,
would indicate interesting physics related to a novel hadron spin
structure \cite{Bur93,Ma99}. Second, the polarization of the
produced $\Lambda$ can be easily determined experimentally by the
reconstruction of its decay products. The self-analyzing property
due to the characteristic decay mode $\Lambda \to p \pi^-$, with a
large branching ratio of 64{\%}, makes it most suitable for
studying the fragmentation of various polarized quarks to
$\Lambda$. In addition, as pointed out by Gribov and
Lipatov \cite{GLR}, the fragmentation function $D_q^h(z)$, for a
quark $q$ splitting into a hadron $h$ with longitudinal momentum
fraction $z$, can be related to the quark distribution $q_h(x)$,
for finding the quark $q$ inside the hadron $h$ carrying a
momentum fraction $x$, by the reciprocity relation
\begin{equation}
D_q^h(z) \sim q_h(x)~.
\end{equation}
$D^h_q$ and $q_h$ depend also on the energy scale $Q^2$ and this
relation holds, in principle, in a certain $Q^2$ range and in
leading order approximation. 
It is important to recall the earlier stronger relation
by Drell, Levy and Yan (DLY) \cite{DLY},
connecting by analytic continuation the DIS structure
functions and the fragmentation functions in $e^+e^-$ collisions.
For a recent
extensive work on the validity of the DLY relation to
$O$$(\alpha_s^2)$, see Ref.~\cite{Blu00}, 
where the Gribov-Lipatov relation
Eq.~(1) is also
verified to hold in leading order for the space- and time-like splitting
functions of QCD.  
Moreover, although Eq.~(1) is only
valid at $x \to 1$ and $z \to 1$, it provides a reasonable
guidance for a phenomenological parametrization  of the various
quark to $\Lambda$ fragmentation functions, since we are still
lacking a good understanding of the spin and flavor structure of
these fragmentation functions. Therefore, at least, we can get
some information on the spin and flavor structure of the $\Lambda$
from its fragmentation functions at large $z$. The flavor symmetry
$SU(3)_F$ in the octet baryons can be also used in order to have a
deeper insight on the nucleon spin structure. 

On the other hand,
there have been some recent progress in the measurements of the
polarized $\Lambda$ production. The longitudinal $\Lambda$
polarization in $e^+e^-$ annihilation at the Z-pole was observed
by several Collaborations at CERN \cite{ALEPH96,DELPHI95,OPAL97}.
Very recently, the HERMES Collaboration at DESY reported the
result of the longitudinal spin transfer to the $\Lambda$ in
polarized positron DIS \cite{HERMES}. Also the E665 Collaboration
at FNAL measured the $\Lambda$ and $\bar{\Lambda}$ spin transfers
from muon DIS \cite{E665}, and they observed very different
behaviour for $\Lambda$ and $\bar{\Lambda}$ polarizations, though
the precision of the data is still rather poor.

Several years ago, Brodsky, Burkardt and Schmidt provided a
reasonable description of the polarized quark distributions of the
nucleon in a pQCD based model \cite{Bro95}. This model has also
been successfully used in order to explain the large single-spin
asymmetries found in semi-inclusive pion production in $pp$
collisions, while other models have not been able to fit the data
\cite{Bog99}. In this paper, we extend this analysis to the
semi-inclusive production of $\Lambda$ and $\bar{\Lambda}$ in DIS
and we try to understand their spin-dependent features.
Especially, a possible sea quark and antiquark asymmetry is tested
against the E665 experimental results in the $\Lambda$ and
$\bar{\Lambda}$ spin transfers.

The paper is organized as follows. In section 2 we will present
various formulae to derive the $z$-dependence
of the spin transfer and polarization of the $\Lambda$
($\bar{\Lambda}$) in lepton DIS. In section 3 we calculate these
spin observables in the pQCD based model, and we find that this
model gives a good description of the available $\Lambda$ data. In
section 4 we present an analysis of the possible contribution from
the sea and we try to outline the different trends of the spin
transfer for $\Lambda$ and $\bar{\Lambda}$ in the E665 experiment.
We test a possible scenario where the sea quarks in the $\Lambda$
( or sea antiquarks in the $\bar{\Lambda}$ ) are negatively
polarized, but the sea antiquarks in the $\Lambda$ ( or sea quarks
in the $\bar{\Lambda}$ ) are positively polarized. Finally, we
present some concluding remarks in section 5.

\section{Spin observables in the $\Lambda$ ($\bar \Lambda$) fragmentation}

There are available data on polarized $\Lambda$ ($\bar{\Lambda}$)
fragmentation functions in $e^+e^-$ annihilation at the Z-pole and
also in lepton DIS. The $\Lambda$ polarization in the $e^+e^-$
annihilation at the Z-pole was previously analysed \cite{MSY3},
and here we concentrate on the spin transfer for the $\Lambda$
production in lepton DIS.
For a longitudinally polarized charged
lepton beam and an unpolarized target, the $\Lambda$ polarization
along its own momentum axis is given in the quark parton model by
\cite{Jaf96}
\begin{equation}
P_{\Lambda}(x,y,z) = P_B D(y)A^{\Lambda}(x,z)~,
\label{PL}
\end{equation}
where $P_B$ is the polarization of the charged lepton beam, which
is of the order of 0.7 or so \cite{HERMES,E665}. $D(y)$, whose
explicit expression is
\begin{equation}
D(y)=\frac{1-(1-y)^2}{1+(1-y)^2},
\end{equation}
is commonly referred to as the longitudinal depolarization factor
of the virtual photon with respect to the parent lepton, and
\begin{equation}
A^{\Lambda}(x,z)= \frac{\sum\limits_{q} e_q^2 [q^N(x,Q^2) \Delta
D_q^\Lambda(z,Q^2) + ( q \rightarrow \bar q)]}
{\sum\limits_{q} e_q^2 [q^N (x,Q^2)
D^\Lambda_q(z,Q^2) + ( q \rightarrow \bar q)]}~,
\label{DL}
\end{equation}
is the longitudinal spin transfer to the $\Lambda$. Here $y=\nu/E$
is the fraction of the incident lepton's energy that is
transferred to the hadronic system by the virtual photon. We see
that the $y$-dependence of $P_{\Lambda}$ factorizes and it can be
reduced to a numerical coefficient when $D(y)$ is integrated over
a given energy range, corresponding to some experimental cuts. In
Eq.~(\ref{DL}), $q^N(x,Q^2)$ is the quark distribution for the
quark $q$ in the nucleon, $ D_q^\Lambda (z,Q^2)$ is the
fragmentation function for $\Lambda$ production from quark $q$,
$\Delta D _q^\Lambda (z, Q^2) $ is the corresponding longitudinal
spin-dependent fragmentation function, and $e_q$ is the quark
charge in units of the elementary charge $e$. In the quark
distribution function, $x=Q^2/{2M \nu}$ is the Bjorken scaling
variable, $q^2=-Q^2$ is the squared four-momentum transfer of the
virtual photon, $M$ is the proton mass, and in the fragmentation
function, $z=E_\Lambda /\nu$ is the energy fraction of the
$\Lambda$, with energy $E_\Lambda$. In a region where $x$ is large
enough, say $0.2 \leq x \leq 0.7$, one can neglect the antiquark
contributions in Eq.~(4) and probe only the valence quarks of the
target nucleon. On the contrary, if $x$ is much smaller, one is
probing the sea quarks and therefore the antiquarks must be
considered as well. 

For $\bar\Lambda$ production the polarization
$P_{\bar\Lambda}$ has an expression similar to Eq.~(2), where the
spin transfer $A^{\bar\Lambda}(x,z)$ is obtained from Eq.~(4) by
replacing $\Lambda$ by $\bar\Lambda$. The $\Lambda$ and
$\bar\Lambda$ fragmentation functions are related since we can
safely assume matter-antimatter symmetry, {\it i.e.}
$D^\Lambda_{q,\bar{q}}(z)=D^{\bar{\Lambda}}_{\bar{q},q}(z)$ and
similarly for $\Delta D^\Lambda_{q,\bar{q}}(z)$. 

It is useful to
consider some kinematics variables in the virtual photon-nucleon
center-of-mass ({\it c.m.}) frame, because the semi-inclusive
$\Lambda$ production in DIS can be also regarded as the inclusive
hadronic reaction $\gamma^* p \rightarrow \Lambda X$. The
experimental results are usually presented as functions
of the variables $x_F$ \cite{E665} or $z'$ \cite{HERMES} rather than
$z$, so we need to make a general kinematics analysis of the
relation of $x_F$ and $z'$ in terms of $z$. From the discussion in
the Appendix we know that $x_F \to z$ and $z' \to z$ in the
Bjorken limit for $z \not= 0$, and also we know that the produced
$\Lambda$ needs to have the same direction as the virtual photon, 
both in the nucleon rest frame and in the {\it c.m.} frame, i.e., the
produced $\Lambda$ is collinear with the virtual photon and it is
in the current fragmentation region. Therefore we can compare the
$z$-dependent predictions of the spin transfers with the
data expressed in terms of $x_F$ or $z'$.

We now turn our attention to the production of any hadron $h$ from
neutrino and antineutrino DIS processes. The longitudinal
polarizations of $h$ in
its momentum direction, for $h$ in the current
fragmentation region can be expressed as,
\begin{equation}
P_\nu^h(x,y,z)=-\frac{[d(x)+\varpi s(x)] \Delta D _u^h (z) -( 1-y)
^2 \bar{u} (x) [\Delta D _{\bar{d}}^h (z)+\varpi \Delta D_{\bar{s}}^h(z)]}
{[d(x)+\varpi s(x)] D_u ^h
(z) + (1-y)^2 \bar{u} (x) [D _{\bar{d}}^h (z)+\varpi D_{\bar{s}}^h(z)]}~,
\end{equation}

\begin{equation}
P_{\bar{\nu}}^h (x,y,z)=-\frac{( 1-y) ^2 u (x) [\Delta D
_d^h (z)+\varpi \Delta D
_s^h (z)]-[\bar{d}(x)+\varpi \bar{s}(x)] \Delta D _{\bar{u}}^h (z)}{(1-y)^2
u (x) [D _d^h (z)+\varpi D
_s^h (z)]+[\bar{d}(x)+\varpi \bar{s}(x)] D_{\bar{u}} ^h (z)}~,
\end{equation}
where the terms with the factor $\varpi=\sin^2 \theta_c/\cos^2
\theta_c$ ($\theta_c$ is the Cabibbo angle) represent Cabibbo
suppressed contributions. The explicit formulae for the case where
$h$ is $\Lambda$ and $\bar{\Lambda}$, including only the Cabibbo
favored contributions, have been given in \cite{Ma99}. We have
neglected the charm contributions both in the target and in hadron
$h$. One advantage of neutrino (antineutrino) process is that the
scattering of a neutrino beam on a hadronic target provides {\it a
source of polarized quarks with specific flavor structure}, and
this particular property makes the neutrino (antineutrino) process
an ideal laboratory to study the flavor-dependence of quark to
hadron fragmentation functions, especially in the polarized case
\cite{Ma99}. The detailed $x$-, $y$-, and $z$- dependencies can
provide more information concerning the various fragmentation
functions. As a special case, the $y$-dependence can be simply
removed by integrating over the appropriate energy range, and the
$x_F$- or $z'$-dependencies of these polarizations can be obtained
using the above formulae. 

It is interesting to notice that, after
consideration of symmetries between different quark to $\Lambda$
and $\bar{\Lambda}$ fragmentation functions \cite{Ma99}, there are
only eight independent fragmentation functions
\begin{equation}
D_q^{\Lambda},
~~ D_{\bar{q}}^{\Lambda},
~~ D_q^{\Lambda}+\varpi D_s^{\Lambda},
~~ D_{\bar{q}}^{\Lambda} +\varpi D_{\bar{s}}^{\Lambda},
\label{uFF}
\end{equation}
and
\begin{equation}
\Delta D_q^{\Lambda},
~~ \Delta D_{\bar{q}}^{\Lambda},
~~ \Delta D_q^{\Lambda}+\varpi \Delta D_s^{\Lambda},
~~ \Delta D_{\bar{q}}^{\Lambda} +\varpi \Delta D_{\bar{s}}^{\Lambda},
\label{pFF}
\end{equation}
where $q$ denotes $u$ and $d$. Different combinations of
unpolarized and polarized $\Lambda$ and $\bar{\Lambda}$
productions in neutrino and antineutrino processes and choices of
specific kinematics regions with different $x$, $y$, and $z$ can
measure the above fragmentation functions efficiently. From
Eqs.~(\ref{uFF}) and (\ref{pFF}) it is possible to extract the
various strange quark fragmentation functions
\begin{equation}
D_s^{\Lambda},
~~ D_{\bar{s}}^{\Lambda},
~~ \Delta D_s^{\Lambda},
~~ \Delta D_{\bar{s}}^{\Lambda},
\end{equation}
provided the accuracy of the data is high enough. This supports
the conclusion of \cite{Ma99} that hadron production in neutrino
(antineutrino) DIS processes provides an ideal laboratory to study
the flavor dependence of the quark fragmentation. Another
advantage of the neutrino (antineutrino) processes is that the
antiquark to $\Lambda$ fragmentation can also be conveniently
extracted, and this can be compared to specific predictions
concerning the antiquark polarizations inside baryons, which in
turn are related to the proton spin problem \cite{Ma99,Bro96}. To
our knowledge, good precision data on $\Lambda$ and
$\bar{\Lambda}$ production will be available soon from the NOMAD
neutrino beam experiment \cite{NOMAD}, thus our knowledge of the
various quark to $\Lambda$ fragmentation functions will be
improved. The theoretical predictions here may provide a practical
guidance for experimental analysis, and a better understanding of
the physics observations.

\section{Description of the $\Lambda$ and $\bar{\Lambda}$
spin observables in a pQCD based model}

In previous works \cite{MSY2,MSY3,MSY4}, the quark
distributions of the $\Lambda$ and other octet baryons at large $x$
have been discussed in the framework of the pQCD based model. In
the region $x \to 1$, pQCD can give
rigorous predictions for the behavior of distribution functions
\cite{Bro95}. In particular, it predicts ``helicity retention",
which means that the helicity of a valence quark will match that
of the parent nucleon. Explicitly, the quark distributions of a
hadron $h$ have been shown to satisfy the counting rule
\cite{countingr},
\begin{equation}
q_h(x) \sim (1-x)^p,
\label{pl}
\end{equation}
where
\begin{equation}
p=2 n-1 +2 \Delta S_z.
\end{equation}
Here $n$ is the minimal number of the spectator quarks, and
$\Delta S_z=|S_z^q-S_z^h|=0$ or $1$ for parallel or anti-parallel
quark and hadron helicities, respectively \cite{Bro95}.

For the $\Lambda$, we have explicit spin distributions for each
valence quark,
\begin{equation}
u^{\uparrow}_v(x)=d^{\uparrow}_v(x)=\frac{1}{x^{\alpha_v}} [
A_{u_v}(1-x)^3+B_{u_v}(1-x)^4 ], 
\label{uvu}
\end{equation}

\begin{equation}
u^{\downarrow}_v(x)=d^{\downarrow}_v(x)=\frac{1}{x^{\alpha_v}} [
C_{u_v}(1-x)^5+D_{u_v}(1-x)^6 ], 
\label{uvd}
\end{equation}

\begin{equation}
s^{\uparrow}_v(x)=\frac{1}{x^{\alpha_v}} [
A_{s_v}(1-x)^3+B_{s_v}(1-x)^4 ], 
\label{svu}
\end{equation}

\begin{equation}
s^{\downarrow}_v(x)=\frac{1}{x^{\alpha_v}} [
C_{s_v}(1-x)^5+D_{s_v}(1-x)^6 ]. 
\label{svd}
\end{equation}
Here $\alpha_v\simeq 1/2 $ is controlled by Regge exchange for
nondiffractive valence quarks. The parameters $A_{u_v}=1.094$,
$B_{u_v}=-0.677$, $C_{u_v}=2.707$, $D_{u_v}=-2.126$,
$A_{s_v}=2.188$, $B_{s_v}=-1.415$, $C_{s_v}=2.707$, and
$D_{s_v}=-2.713$ are taken the same values as in previous paper
\cite{MSY3}, with the valence quark helicities $\Delta s =0.7$ and
$\Delta u =-0.1$. These are slightly different from the
Burkardt-Jaffe values \cite{Bur93} $\Delta s= 0.6$ and $\Delta u=
-0.2$, to reflect the fact that the sea quarks might contribute
partially to the total $\Delta s$ and $\Delta u$. With these set
of parameters, we can have a good description of the $\Lambda$
polarization in the $e^+ e^-$ annihilation process at the Z-pole
\cite{MSY3}. 

The calculated $z$-dependence of the $\Lambda$ and
$\bar{\Lambda}$ polarizations in lepton DIS process with $x$
integrated over the range [0.02, 0.4] are presented, as dotted
curves, in Fig.~\ref{mssy5f1} and Fig.~\ref{mssy5f2} respectively.
In our numerical calculations, the CTEQ5 parton distributions of
the nucleon at $Q^2=1.3 GeV ^2$ are adopted \cite{CTEQ5}. We find
that our prediction of the $z$-dependence of the $\Lambda$
polarization in Fig.~\ref{mssy5f1} is consistent with the HERMES
experimental data \cite{HERMES}. The data seem to indicate a trend
supporting the prediction \cite{MSY2,MSY3} of positively polarized
$u$ and $d$ quarks inside $\Lambda$ at large $x$. Our prediction
is also in qualitative agreement with a Monte Carlo simulation
based on inputs of the naive quark model and a model with SU(3)
symmetry \cite{Ash99}. The $z$-dependence of the fragmentation
functions in the Monte Carlo simulation has more validity than
naive assumptions without $z$-dependence \cite{Kot98}, and this
supports our $z$-dependence predictions of the quark to $\Lambda$
fragmentation functions based on physics arguments
\cite{MSY2,MSY3}.
With the same parton distributions, the $\Lambda$ and $\bar{\Lambda}$
polarizations for the neutrino and antineutrino DIS processes are
predicted, as shown, as dotted curves, in Figs.~\ref{mssy5f5} -
\ref{mssy5f8}. We have also compared the situations with and
without the Cabibbo suppressed contributions, but we found that
the modifications due to the Cabibbo suppression are so small that
they can be ignored. However, this small modification from the
Cabibbo suppressed terms is specific to the pQCD based model and
is not true in general for such situations with very small
$D_q^{\Lambda}/D_s^{\Lambda}$ and $\Delta D_q^{\Lambda}/\Delta
D_s^{\Lambda}$. For example, the contributions from the Cabibbo
suppressed terms will be dominant in the situation with $\Delta
D_q^{\Lambda}=0$. We also point out here that our predictions are
dramatically different from the previous calculation \cite{Kot98}
based on the assumptions 
$\Delta D_q^{\Lambda}(z)= C_q^{\Lambda} D_q^{\Lambda}(z)$
where $C_q^{\Lambda}$ is a constant or a slowly varying function of $z$. 
In our case $C_q^{\Lambda}$ is a function of $z$
and it even changes sign for $q=u,d$ when $z$ varies between 0 and 1
\cite{MSY2,MSY3}. The behaviour near $z=1$, which contrasts
with that of Ref.~\cite{Kot98}, is directly
related to our explicit parametrization  of 
the quark distributions (see Eqs.~(\ref{uvu}, \ref{uvd}) ).


\begin{figure}
\begin{center}
\leavevmode {\epsfysize=8cm \epsffile{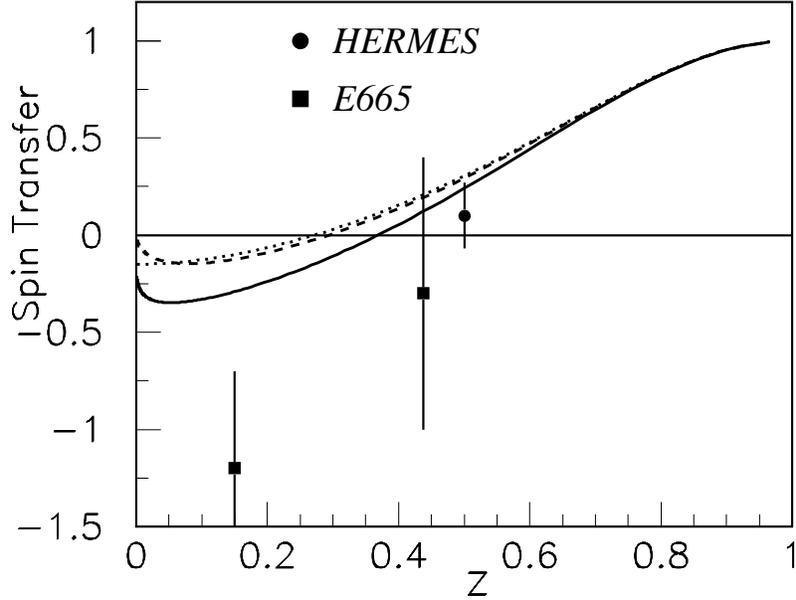}}
\end{center}
\caption[*]{\baselineskip 13pt The $z$-dependence of the $\Lambda$
spin transfer in electron or positron (muon) DIS. The dotted curve
corresponds to pure valence quark contributions. The solid and
dashed curves are with the contributions of sea quarks for
scenario I and scenario II, respectively (see section 4). Note
that for HERMES data the $\Lambda$ polarization is measured along
the virtual-photon momentum, whereas for E665 it is measured along
the virtual-photon spin. The averaged value of the Bjorken
variable is chosen as $x=0.1$ (corresponding to the HERMES
averaged value) and the calculated result is not sensitive to a
different choice of $x$ in the small $x$ region (for example,
$x=0.005$ corresponding to the E665 averaged value).
}\label{mssy5f1}
\end{figure}


\begin{figure}
\begin{center}
\leavevmode {\epsfysize=8cm \epsffile{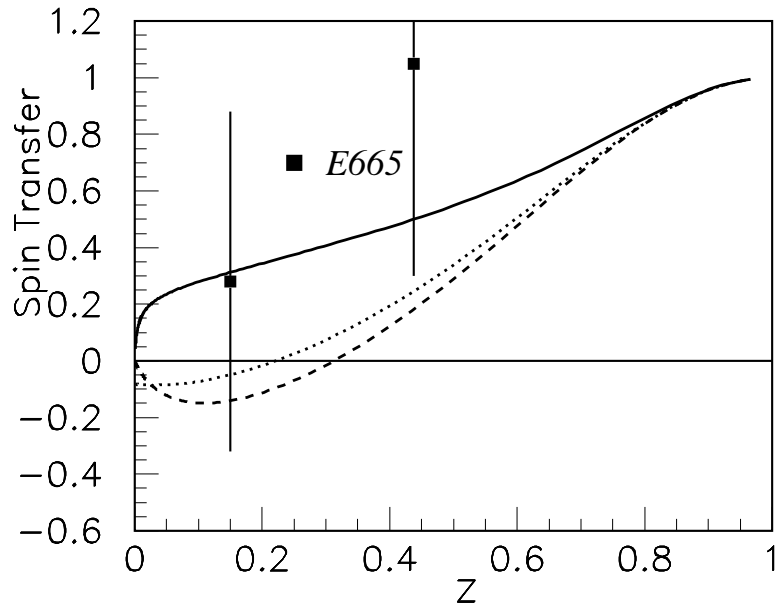}}
\end{center}
\caption[*]{\baselineskip 13pt The $z$-dependence of the
$\bar{\Lambda}$ spin transfer in electron or positron (muon) DIS.
The dotted curve corresponds to pure valence quark contributions.
The solid and dashed curves are with the contributions of sea
quarks for scenario I and scenario II, respectively (see section
4). }\label{mssy5f2}
\end{figure}


\begin{figure}
\begin{center}
\leavevmode {\epsfysize=12cm \epsffile{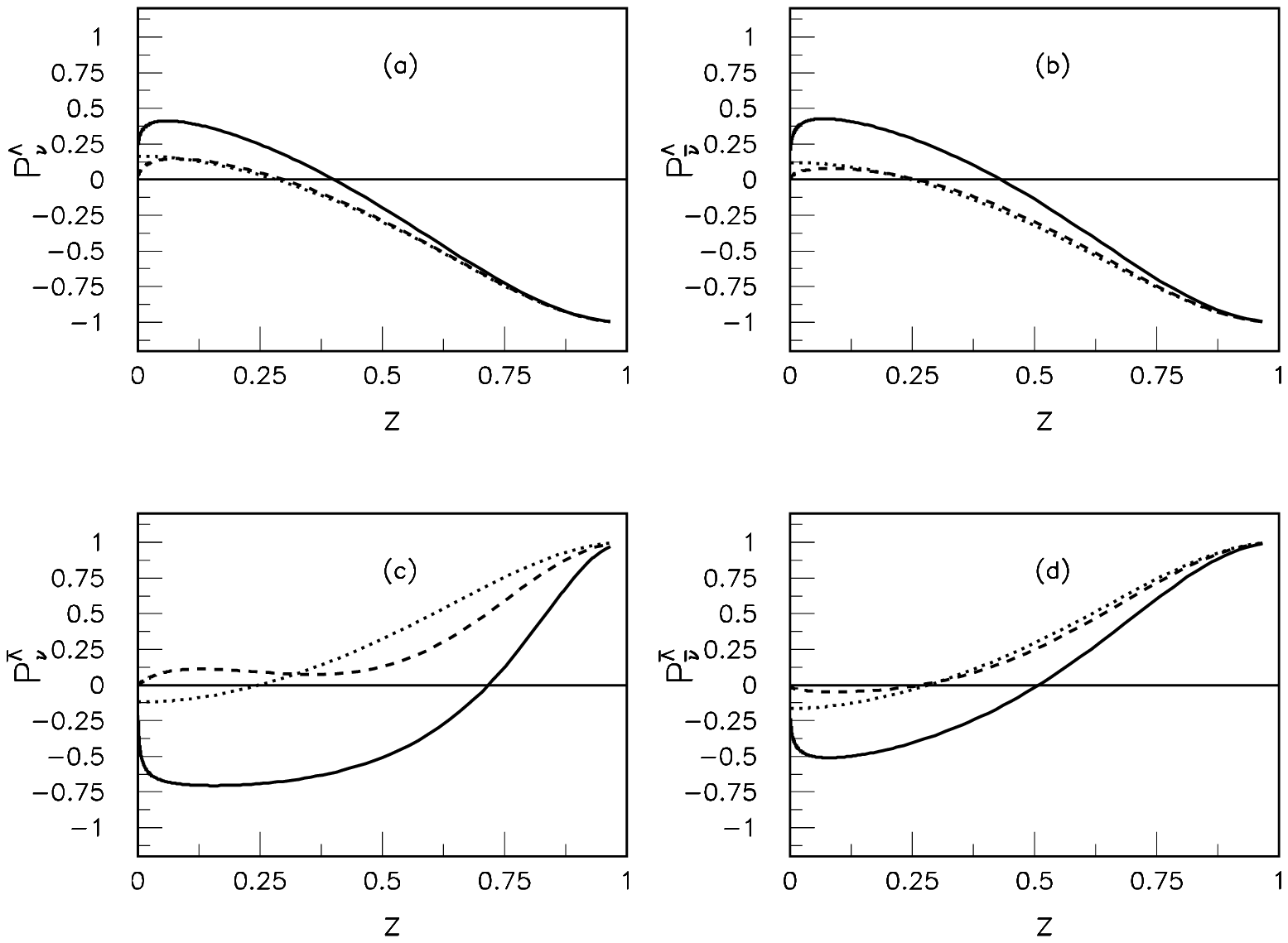}}
\end{center}
\caption[*]{\baselineskip 13pt The $z$-dependence of the $\Lambda$
and $\bar{\Lambda}$ polarizations for the neutrino (antineutrino)
DIS processes. The dotted curve corresponds to pure valence quark
contributions. The solid and dashed curves are with the
contributions of sea quarks for scenario I and scenario II,
respectively (see section 4). Note that the dashed and dotted
curves in (a) and (b) almost overlap. $x$ and $y$ are integrated
in the ranges [0.02, 0.4] and [0, 1], respectively.
}\label{mssy5f5}
\end{figure}


\begin{figure}
\begin{center}
\leavevmode {\epsfysize=12cm \epsffile{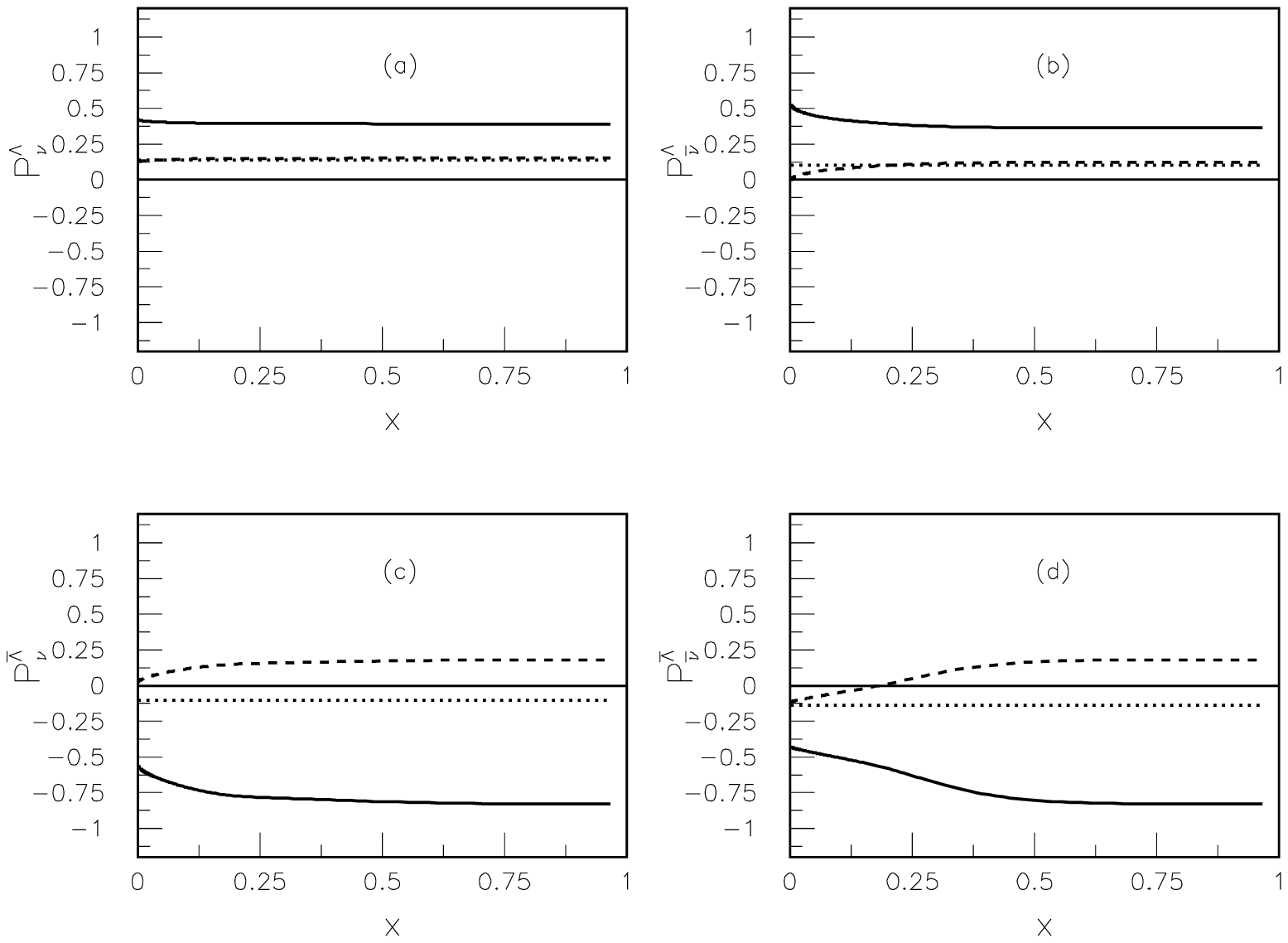}}
\end{center}
\caption[*]{\baselineskip 13pt The $x$-dependence of the $\Lambda$
and $\bar{\Lambda}$ polarizations for the neutrino (antineutrino)
DIS processes at $z=0.1$. The dotted curve corresponds to pure
valence quark contributions. The solid and dashed curves are with
the contributions of sea quarks for scenario I and scenario II,
respectively (see section 4). Note that the dashed and dotted
curves in (a) and (b) almost overlap. $y$ is integrated in the
range [0, 1]. }\label{mssy5f7}
\end{figure}


\begin{figure}
\begin{center}
\leavevmode {\epsfysize=12cm \epsffile{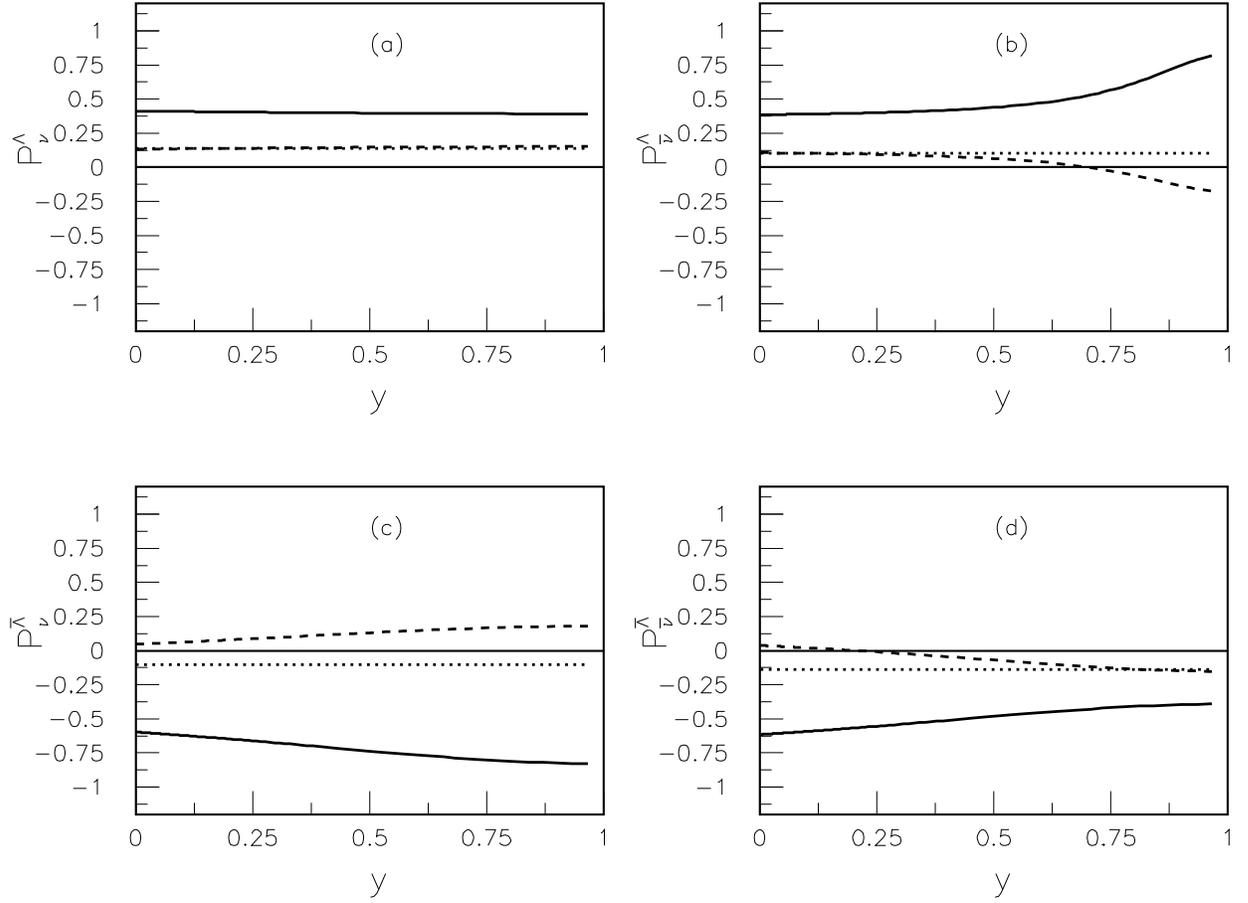}}
\end{center}
\caption[*]{\baselineskip 13pt The $y$-dependence of the $\Lambda$
and $\bar{\Lambda}$ polarizations for the neutrino (antineutrino)
DIS processes at $z=0.1$ and $x=0.1$. The dotted curve corresponds
to pure valence quark contributions. The solid and dashed curves
are with the contributions of sea quarks for scenario I and
scenario II, respectively (see section 4). Note that the dashed
and dotted curves in (a) almost overlap. }\label{mssy5f8}
\end{figure}

\newpage

\section{Possible sea quark contributions to the $\Lambda$ and $\bar{\Lambda}$ spin
observables} 

We now look at the sea quark contributions to the
$\Lambda$ polarization in order to find the rough shapes of the
sea quark and sea antiquark in the pQCD based model. Strictly
speaking, the Gribov-Lipatov relation Eq.~(1) has limitations for
its application at small $x$ \cite{Pet99}, therefore we should
consider our method as a search for a reasonable parametrization 
of quark to $\Lambda$ fragmentation functions, and then check the
validity and relevance of the parametrization  by comparing
predictions with experimental measurements of various processes.
Also there is
still a large freedom in the detailed treatments and many
assumptions are needed, so that the predictive power at small $z$
has more limitations than that at large $z$ in our analysis.
However, the method has been supported by comparison of the prediction with
the experimental data
of $\Lambda$ polarizations from both $e^+e^-$-annihilation at the $Z$-pole
\cite{MSY3} and polarized lepton on the nucleon target
DIS scattering \cite{MSY2}, and
the uncertainties can be gradually constrained and
reduced with more data later on.

The sea quark
helicity distributions in the pQCD analysis \cite{Bro95} satisfy
\begin{equation}
q_s^{\uparrow}=\frac{1}{x^{\alpha_s}} [
A_{q_s}(1-x)^5+B_{q_s}(1-x)^6 ], \label{qsu}
\end{equation}

\begin{equation}
q_s^{\downarrow}=\frac{1}{x^{\alpha_s}} [
C_{q_s}(1-x)^7+D_{q_s}(1-x)^8 ], \label{qsd}
\end{equation}
\begin{equation}
\bar{q}^{\uparrow}=\frac{1}{x^{\alpha_s}} [
A_{\bar{q}}(1-x)^5+B_{\bar{q}}(1-x)^6 ], \label{qsbu}
\end{equation}

\begin{equation}
\bar{q}^{\downarrow}=\frac{1}{x^{\alpha_s}} [
C_{\bar{q}}(1-x)^7+D_{\bar{q}}(1-x)^8 ],
\label{qsbd}
\end{equation}
where $\alpha_s$, which is controlled by Regge exchange for sea
quarks, is taken as the same as that in Ref.\cite{Bro95}, i.e.
$\alpha_s\simeq 1.12 $. We also 
take the same $q_s^{\uparrow,\downarrow}$
and $\bar{q}^{\uparrow,\downarrow}$ for $q=u,d,s$, for the sake of simplicity. 
We constrain the sea quark distributions
by three conditions:\\ i)~$A_q+B_q=C_q+D_q$ for $q=q_s$ and
$\bar{q}$ from the convergence of sum rules \cite{Bro95};\\
ii)~the values of $\Delta q_s$ and $\Delta \bar{q}$;\\ iii)~the
momentum fractions carried by sea quarks $\left<x_{q_s}\right>$
and sea antiquarks $\left<x_{\bar{q}}\right>$.\\ This leaves us
with one unknown parameter for each set of $A_q$, $B_q$, $C_q$,
and $D_q$. In the following discussions, the values of $\Delta
q_s$ and $\Delta \bar{q}$ are taken so as to let $\Delta U =
\Delta u_v + \Delta q_s +\Delta \bar{q}=-0.2$ and $\Delta S =
\Delta s_v + \Delta q_s +\Delta \bar{q}=0.6$ be consistent with
the values of Burkardt-Jaffe sum rule results \cite{Bur93}. 

It has
been pointed out in Ref.~\cite{Bro96} that there might be
quark-antiquark asymmetry in the quark-antiquark pairs of the
nucleon sea, and a possibility to check the strange
quark-antiquark asymmetry of the nucleon sea through strange quark
and antiquark fragmentations to proton has been also suggested in
Ref.~\cite{Bro97}. Therefore we can consider the possibility of
sea quark-antiquark asymmetry in the $\Lambda$. For comparison we
introduce two scenarios, with asymmetric quark-antiquark sea as
scenario I, and symmetric quark-antiquark sea as scenario II. 

For
scenario I we choose asymmetric quark-antiquark helicity sums
$\Delta q_s=-0.3$ and $\Delta \bar{q}=0.2$ with
$\left<x_{q_s}\right>=\left<x_{\bar{q}}\right> \approx 0.03$. We
choose $C_{q_s}$ as the free parameter and other three parameters
are given by the three constrains mentioned before as the solution
of
\begin{equation}
\begin{array}{clllc}
A_{q_s}=~~0.988 C_{q_s}-1.692
\\
B_{q_s}=-1.117 C_{q_s}+1.915
\\
D_{q_s}=-1.129 C_{q_s}+0.222
\end{array}
\label{abdqs}
\end{equation}
The probabilistic interpretation of parton distributions
$q_s^{\uparrow}$ and $q_s^{\downarrow}$ implies the rather
stringent bounds
\begin{equation}
1.714 < C_{q_s} < 1.718.
\end{equation}
Similarly, the parameters for the sea antiquarks are constrained
by
\begin{equation}
\begin{array}{clllc}
A_{\bar{q}}=~~0.988 C_{\bar{q}} + 0.860
\\
B_{\bar{q}}=-1.117 C_{\bar{q}} - 0.846
\\
D_{\bar{q}}=-1.129 C_{\bar{q}} + 0.0143
\end{array}
\label{abdqs2}
\end{equation}
with
\begin{equation}
0 < C_{\bar{q}} < 0.111.
\end{equation}

In the following calculation, we take $C_{q_s}=1.715$ and
$C_{\bar{q}}=0.1$. From the solid curves in Figs.~\ref{mssy5f1}
and \ref{mssy5f2}, we find that the pQCD based model with sea
contributions can reproduce the different behaviors of the
$\Lambda$ and $\bar{\Lambda}$ spin transfers, as observed in the
E665 experimental data \cite{E665}. This implies that the
different behaviors with quark-antiquark asymmetry of sea quark
fragmentations might be a source for the $\Lambda$ and
$\bar{\Lambda}$ spin transfers difference observed by the E665
collaboration. Needless to mention that, although the E665 data is
of poor precision, the different behaviors of the $\Lambda$ and
$\bar{\Lambda}$ spin transfers might still be a genuine effect.
The magnitude of the measured spin transfer Eq.~(\ref{DL}) should
be less than unity and also $x_F \approx z$ is
a good approximation in the kinematics range of the E665
experiment. Also Fig.~\ref{mssy5f5} indicates that there is a big
modification to the neutrino induced $\bar{\Lambda}$ polarization
in the small and medium $z$ region with the sea contributions
included, thus $\Lambda$ and $\bar{\Lambda}$ production in
neutrino (antineutrino) DIS processes may provide relevant
information concerning the antiquark to $\Lambda$ fragmentation
functions. However, as has been discussed in \cite{Bro96}, the
antiquarks inside the baryons are likely to be unpolarized or
slightly positive polarized from a baryon-meson fluctuation model
of intrinsic sea quark-antiquark pairs. Therefore we expect that
the $\Lambda$ and $\bar{\Lambda}$ productions from neutrino
(antineutrino) processes will provide more information about
the antiquark polarizations inside baryons, or more precisely, 
the antiquark to $\Lambda$ fragmentation functions. 

To
reflect the role played by the quark-antiquark asymmetry in
reproducing the different behaviors for the fragmentations of
$\Lambda$ and $\bar{\Lambda}$ in electron or positron (muon) DIS
processes, we introduce scenario II of symmetric quark-antiquark
helicity sums $\Delta q_s=\Delta \bar{q}=-0.05$ with
$\left<x_{q_s}\right>=\left<x_{\bar{q}}\right> \approx 0.03$. We
have

\begin{equation}
\begin{array}{clllc}
A_{q_s}=A_{\bar{q}}=~~~0.988 C_{q}-0.416
\\
B_{q_s}=B_{\bar{q}}=-1.117 C_{q}+0.534
\\
D_{q_s}=D_{\bar{q}}=-1.129 C_{q}+0.118
\end{array}
\end{equation}
with
\begin{equation}
0.422< C_{q} < 0.914.
\end{equation}
The calculated results with $C_{q}=0.6$ (for both $q=q_s$ and
$\bar{q}$) are also presented in Figs.~\ref{mssy5f1} and
\ref{mssy5f2}. From Figs.~\ref{mssy5f1} and \ref{mssy5f2} we find
that in the scenario II we can only produce a small difference of
$\Lambda$ and $\bar{\Lambda}$ fragmentations in electron or
positron (muon) DIS process. This shows the necessity of having an
asymmetric quark-antiquark to $\Lambda$ fragmentations as a
possibility to understand the different behaviours of the
$\Lambda$ and the $\bar{\Lambda}$ spin transfers in the E665
experiment \cite{E665}. We also present in
Figs.~\ref{mssy5f5}-\ref{mssy5f8} the results from scenario II and
notice the different predictions which can be tested in neutrino
(antineutrino) DIS processes.

\section{Concluding remarks}

In summary, we investigated the $\Lambda$ and $\bar{\Lambda}$
polarizations in lepton DIS in the pQCD based model. We find that
the model can give a good description of the available data for
the spin transfer. The $\Lambda$ and $\bar{\Lambda}$ polarizations
in the neutrino DIS process are predicted. We find that sea
contribution gives a big modification to the spin transfer in the
small $z$ region. The E665 experimental data show very different
behaviors of the $\Lambda$ and $\bar{\Lambda}$ spin transfers
\cite{E665}, which suggests a possible situation such that the sea
quarks in the $\Lambda$ ( or sea antiquarks in the
$\bar{\Lambda}$) are large negatively polarized, but the sea
antiquarks in the $\Lambda$ ( or sea quarks in the $\bar{\Lambda}$) 
are positively polarized.

{\bf Acknowledgments: } We are very grateful to A.~Kotzinian for his
useful discussion on the relation between $z$ and $x_F$.
This work is partially supported by
Fondecyt (Chile) postdoctoral fellowship 3990048, by the
cooperation programmes Ecos-Conicyt and CNRS- Conicyt between
France and Chile, by Fondecyt (Chile) grant 1990806 and by a
C\'atedra Presidencial (Chile), and by National Natural Science
Foundation of China under Grant Numbers 19605006, 19875024,
19775051, and 19975052.

\appendix

\setcounter{equation}{0}
\renewcommand\theequation{A.\arabic{equation}}
\section*{Appendix: a general kinematics analysis}
We discuss here the kinematics of the semi-inclusive 
hadron production process from a virtual boson scattering 
on a hadronic target. For the physics consideration, 
we take the hadron as
$\Lambda$, the boson as photon, and the hadronic 
target as nucleon, but the discussion 
applies also
to the case 
of any other hadron, boson, and target.
For the general validity of the analysis, we consider
the production of the outgoing hadron and the hadronic debris X, 
treated as an effective particle, regardless 
of the related sub-process in terms of quarks and gluons. 

We first consider the kinematics of the particles in the target
rest frame. The four-momenta of the incident virtual photon and
the nucleon target are
\begin{equation}
q = (\nu,{\mathbf{q}}),
~~~~~~p=(M,{\mathbf{0}}),
~~~~~{\mathrm{with}}~~~~ Q^2={\mathbf{q}}^2-\nu^2~,
\end{equation}
and those of the outgoing $\Lambda$ and the hadronic debris
with effective mass $M_X$ and momentum ${\mathbf{p}}_X$
are
\begin{equation}
p_{\Lambda}=(E_{\Lambda}, {\mathbf{p}}_{\Lambda}),  
~~~~~~p_X=(E_X,{\mathbf{p}}_X)
~~~{\mathrm{with}}~~~{\mathbf{p}}_X={\mathbf{q}}-{\mathbf{p}}_{\Lambda}.
\end{equation}
Using the overall energy conservation we get
\begin{equation}
M_X^2=(\nu+M-E_{\Lambda})^2-{\mathbf{p}}^2_X.
\end{equation}
When expressed in terms of $x=Q^2/2M\nu$, 
$z=E_{\Lambda}/\nu$, and $\nu$, we obtain
\begin{equation}
M_X^2=2M\nu (1-x-z)- 2 z \nu ^2 
\left[1-\sqrt{1+\frac{2M x}{\nu}}\sqrt{1-\frac{M^2_{\Lambda}}{z^2\nu^2}}
\cos \theta \right] +M^2+M^2_{\Lambda},
\label{MX}
\end{equation}
where $\theta$ is the angle of ${\mathbf{p}}_{\Lambda}$ relative
to ${\mathbf{q}}$. 

We now consider the kinematics for the particles in the virtual
photon and target nucleon {\it c.m.} frame where we have
\begin{equation}
q'=(\nu',{\mathbf{q}}')~,
~~p'=(E',-{\mathbf{q}}')~,
~~p'_{\Lambda}=(E'_{\Lambda},{\mathbf{p}}'_{\Lambda})~,
~~p'_X=(E'_X,-{\mathbf{p}}'_{\Lambda})~.
\end{equation}
We recall that the total energy square of the ``two-body" 
reaction $\gamma^* p \rightarrow \Lambda X$ is
\begin{equation}
W^2=(p+q)^2=-Q^2+2 M \nu +M ^2= 2 M \nu (1-x )+M^2.
\end{equation}
Standard two-body kinematics gives
\begin{equation}
E'=\frac{W^2+M^2+Q^2}{2 W}=\frac{M \nu +M^2}{W}
\label{E'}
\end{equation} 
and
\begin{equation}
E'_{\Lambda}=\frac{W^2+M^2_{\Lambda} -M^2_X}{2W}~.
\label{E'Lambda}
\end{equation} 
Now from
\begin{equation}
p \cdot p_{\Lambda}=M E_{\Lambda}= p' \cdot p'_{\Lambda}
=E' E'_{\Lambda}+ {\mathbf{q}}'\cdot {\mathbf{p}}'_{\Lambda},
\end{equation}
we have
\begin{equation}
\cos \theta_{CM}=\frac{M E_{\Lambda}-E' E'_{\Lambda}}{|{\mathbf{q'}}| |{\mathbf{p'}}_{\Lambda}|}~,
\label{theta2}
\end{equation}
where $\theta_{CM}$ is the angle of ${\mathbf{p'}}_{\Lambda}$ relative
to ${\mathbf{q'}}$. 
As we have mentioned in section 2, in the experimental measurements
one usually introduces 
two variables attached to the produced
particle $\Lambda$: the first one is the Feynman variable $x_F = 2
p'_L/W$, where $p'_L = |{\mathbf{p'}}_{\Lambda}|\cos \theta_{CM}$
and the second one is the variable $z'$ defined as
\begin{equation}
z^\prime = \frac{E_\Lambda^\prime}{E' (1-x)}~.
\label{zp}
\end{equation}
 From the above kinematics analysis, we can express $x_F$ and $z'$
in terms of $z$, $x$, $\nu$, and $\theta$. 
In the Bjorken limit $ \nu \to \infty$ with $0 < x <1$,  
we have from (\ref{MX}), for $z \not=0$ and $\theta \not =0$,
\begin{equation}
M^2_X \to -2 z \nu^2 (1-\cos \theta),
\end{equation} 
which leads to a negative energy $E'_X \to -z \nu^2 (1-\cos \theta)/W$
and is unphysical for $\theta \not= 0$. 
Therefore we must have $\theta =0 $ and in this case, the Bjorken limit
leads to
\begin{equation}
M^2_X \to 2 M \nu (1-x) (1-z),
\end{equation}
and from (\ref{E'}) and (\ref{E'Lambda}) we have
\begin{equation}
|{\mathbf{q'}}| \sim E'\rightarrow M \nu/W ~~~~ {\mathrm{and}} ~~~
|{\mathbf{p'}}_{\Lambda}| \sim E'_{\Lambda} \rightarrow M\nu (1-x)z/W ~.
\end{equation}
 From Eq.~(\ref{theta2}), we find that
\begin{equation}
\cos \theta_{CM} \to 1,
\end{equation}
which implies that the produced $\Lambda$ (or $\bar{\Lambda}$) 
is also along the virtual photon direction (i.e., $\theta_{CM}=0$)
in the the {\it c.m.} frame. This corresponds to the current fragmentation
region since $x_F >0$.

Finally from the definitions of $x_F$ and $z$, by using (\ref{MX}), (\ref{E'})
and (\ref{E'Lambda}), we immediately find that
\begin{equation}
x_F \to z ~~~ {\mathrm{and}} ~~~ z' \to z.
\end{equation}

\end{document}